\documentstyle[11pt,newpasp,twoside,epsf]{article}
\def\edcomment#1{\iffalse\marginpar{\raggedright\sl#1\/}\else\relax\fi}
\reversemarginpar

\begin{document}
\title{NIR and Optical Structural Parameters of Galaxies 
in the Cluster AC118 at z=0.31}

\author{G. Busarello, P. Merluzzi, M. Massarotti}
\affil{Osservatorio Astronomico di Capodimonte, Napoli, Italy}
\author{F. La Barbera, M. Capaccioli}
\affil{Physics Department, Universit\`a Federico II, Napoli, Italy}
\author{G. Theureau}
\affil{Observatoire de Paris, Paris, France}

\begin{abstract}
We present some preliminary results related to a project aimed at
studying the evolution of the galaxy population in rich environments by
means of the Color-Magnitude relation and of the Fundamental Plane. We
derive the NIR and optical structural parameters for a sample of galaxies
in the cluster AC118 at z=0.31. We prove that reliable structural parameters 
of galaxies at z$\sim$0.3 can be still derived from ground--based 
observations.
The NIR effective radii, measured for the first
time at this redshift, turn out to be significantly smaller 
than  those derived from the optical data, providing new
insight into the evolution of colour gradients in galaxies.\footnote
{Based on observations at ESO NTT 
(OAC guaranteed time) and HST Data Archive.} 
\end{abstract}

\section{Introduction}

Numerous photometric surveys are
currently done on wide areas of the sky with ground-based telescopes.  
They are providing an enormous
output of data that, in particular, are quickly improving our knowledge
of distant galaxies. 
In addition to integrated quantities like magnitudes and colours, the
possibility of extracting galaxy structural information would significantly
increase their scientific outcome.

Structural parameters have been recently derived for distant galaxies from
HST photometry in optical wave bands (Kelson et al. 2000, and references therein).
J\o rgensen et al. (1999) proved that structural parameters can be still 
derived from ground--based data up to z=0.18. We now extend the redshift range 
and present the first measurement of NIR effective radii at z=0.31.
\vspace{-0.1cm}

\begin{figure}[!h]
\plotone{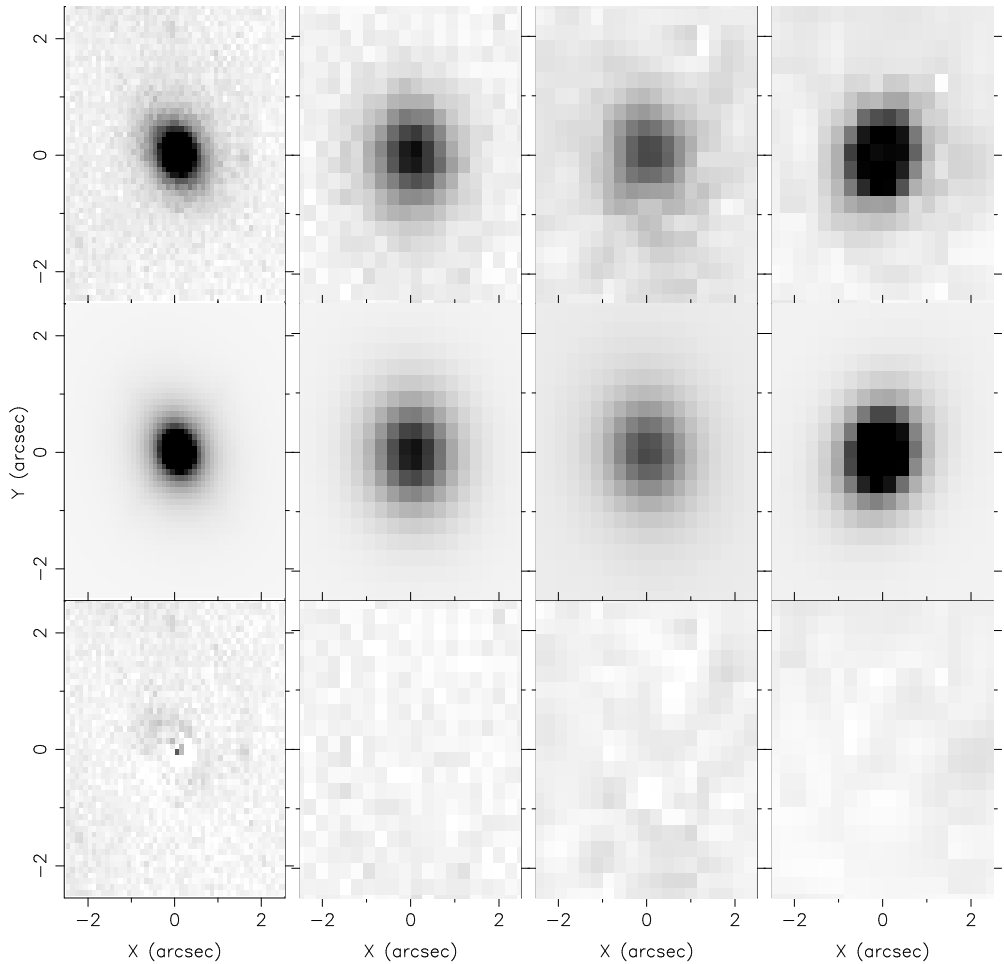}
\caption{{\it Upper panel:} image of a galaxy (from left to right:
HST--F702W, NTT R, V, and K). {\it Middle panel:} model
used to fit the galaxy surface brightness distribution. {\it Bottom panel:}
residuals of the fit. The galaxy orientation changes in the different 
wave bands.}
\end{figure}

\section{The Data}

We obtained photometry in the V, R, I, K bands for the cluster of 
galaxies AC118 with the ESO NTT telescope (EMMI and SOFI) during four
observing runs (October 1998 -- September 2000).  Here we will make use of
the optical and K-band images, and HST (WFPC2-F702W) archive data to derive 
surface photometry for a sample of cluster members.  

The data reduction and photometric calibration were performed following
the standard procedures and are described elsewhere. The sample of
cluster members were selected via photometric redshifts.

\begin{figure}[!h]
\plotone{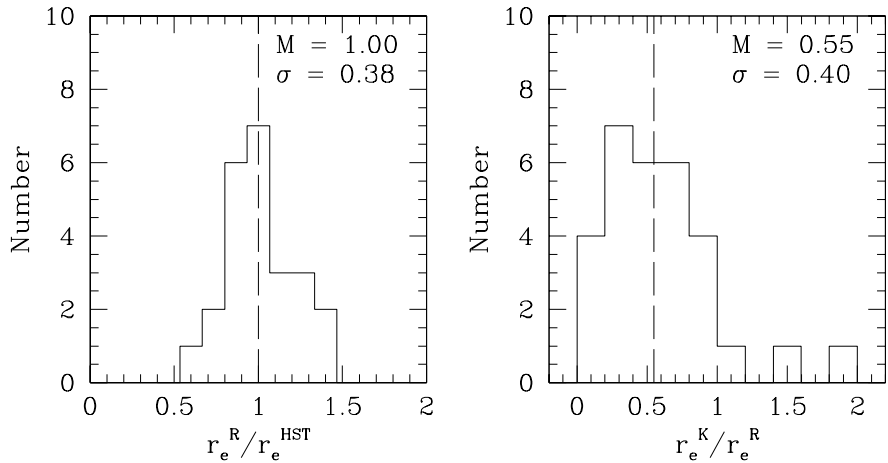}
\caption{{\it Left panel}: comparison between the effective radii
estimated from ground--based (R) and HST (F702W) data. {\it Right
panel}: comparison between the effective radii estimated from R and K
data. The median and standard deviation of the distributions are
indicated.}
\end{figure}

\section{Structural Parameters}

Galaxy sizes at intermediate redshifts are comparable with the typical
seeing of ground-based images ($\sim 1''$). To obtain reliable estimates
of the galaxy structural parameters, it is therefore crucial to take into 
account the effects of the Point Spread Function (PSF).

This is usually performed 1) by a one dimensional fitting approach
(see Saglia et al. 1997) and 2) by two--dimensional fitting methods (see
e.g. van Dokkum \& Franx 1996).  In case 1) the integrated light curve
or the galaxy intensity profile are constructed by an elliptical
fit of the galaxy isophotes and fitted with a proper model.
Due to the small galaxy size in ground-based images, the 
isophotal fit has to be done with few pixels and the brightness profiles 
consist of very few data points.  In the two--dimensional
approach, a direct fit of the galaxy brightness distribution is
performed without any intermediate step. Because of these reasons, we
adopted the two-dimensional approach (see Figure 1).

\section{Optical and NIR Effective Radii}

Structural parameters were derived in the optical wave bands for galaxies
brighter than R=21 mag. The effective radii derived from the HST (F702W)
and from the NTT (R) images are compared in Figure 2 (left panel).
The fitted parameters are fully consistent, with a dispersion of $\sim$40\%,
that is the typical uncertainty on effective radii measurements (see
Kelson et al. 2000). This result shows that reliable structural parameters
of galaxies can be still derived from ground--based observations at z$\sim$0.3.

The right panel of Figure 2 shows the ratios of the NIR to the optical 
effective radii. The K--band (rest-frame H) effective radii turn out
to be $\sim$50\% smaller than those measured in the R--band
(rest-frame V). This implies the presence of strong internal (optical-NIR) 
colour gradients in galaxies at z=0.31. Following Sparks \& J\o rgensen 
(1993, eq. 21) we obtain $\Delta$(R-K)/$\Delta$log(r)$\sim$-0.28 at z=0.31.
When compared with the local (z$\sim$0) value  
$\Delta$(V-H)/$\Delta$log(r)$\sim$-0.18 (Scodeggio et al. 1998, see 
also Peletier et al. 1990), this result implies that the colour gradients 
of galaxies decreased by a factor $\sim$1.5 since z=0.31 (or $\sim$4.5 Gyr).
In a forthcoming paper we will make a more reliable comparison between optical
and NIR structural parameters by extending the sample to $\sim$100 cluster 
galaxies.

\end{document}